\documentstyle[11pt,epsfig]{article}
\begin{document}
\centerline {\large {\bf {Dynamic Work Distribution for PM Algorithm}}}
\vskip 0.5cm

\centerline {Ettore Carretti,$^{(1)}$ Antonio Messina$^{(2)}$}
\vskip 0.5cm

\medskip
\noindent
(1) {\it {I.Te.S.R.E./CNR, Via Gobetti 101, I-40129, Bologna, ITALY}}\\
(2) {\it {Dipartimento di Scienze dell'Informazione, Via Mura
          Anteo Zamboni 7, I-40126, Bologna, ITALY}}\\
{\it e-mail:} carretti@tesre.bo.cnr.it, messina@cs.unibo.it
\vskip 1.0cm

\centerline {\bf {Abstract}}

{\small  Although poor for small dynamic scales, the Particle--Mesh (PM) model
  allows in astrophysics good insight for large dynamic scales at low 
  computational cost. Furthermore, 
  it is possible to employ a very high number of particles to get high 
  mass resolution. 
  These properties could be exploited by suitable parallelization of the 
  algorithm. 

  In PM there are two types of data: the particle data, i.e. position and 
  velocity, which are stored in one--dimensional arrays of $N$ elements, 
  and the grid data, i.e. density and force, which are stored in 
  three--dimensional arrays $M\times M\times M$ in size.  Since 
  individual particles can change cell under the action of gravitational 
  force, parallelization is a real challenge on parallel machine and 
  must account for the machine architecture. 

  We have implemented a dynamic work distribution through agenda parallelism.
  By subdividing the work in small tasks, 
  the implementation is well balanced, scalable and efficient also for 
  clustered particle distributions.

  In this contribution we describe this efficient, load balanced, 
  parallel implementation of PM algorithm on Cray T3E at CINECA 
  and show its performances on cosmological simulation results.
}

\section{Introduction}

N--body simulation codes are powerfull tools to study
the non--linear evolution of gravitational systems.
They are used in cosmology to study the evolution of
the large scale structure of the universe and formation
and evolution of galaxy clusters.

Starting from linear initial condition described by
cosmological models, N--body codes provide the actual distribution of
a N-particle system due only to gravitational interaction.

In the study of the large-scale structure it is important to perform simulations
with a high number of particles, because higher mass resolution can be achieve.

Being the fastest of all known algorithms, the PM algorithm 
(Hockney \& Eastwood, 1981; Birdsall \& Langdom, 1985) can be effectively
used to study the cluster and large-scale evolutions, where we want
to consider the contributes of the very-large scales in the universe.

At the moment, massive parallel processors (MPPs) are the most powerful
computers available. Implementing parallel PM algorithm is thus needed in order
to make simulations with as high as possible number of particles
(Ferrell \& Bertschinger, 1994; Pearce \& Couchman, 1997;
MacFarland et al., 1998; Carretti, 1999).

MPP machines offer very different architectures and require different
parallelization strategies in order to achieve high efficiency.

We have performed a dynamic parallel implementation of the PM algorithm on the
Cray T3E at CINECA Computing Center. We have used the Cray
proprietary {\em shmem} library, which performs message-passing among
processors by fully exploiting the machine hardware.

\section {PM Algorithm} 

The PM algorithm solves the gravity field equations

\begin{equation} 
   \nabla^2\phi = 4\pi G\rho\, , \label{eq01}
\end{equation}

\begin{equation}
   {\bf{F}} = - {\bf \nabla} \phi \, , \label{eq02}
\end{equation}

\noindent
on a grid.
We will call $N$ the particle number and $M$ the one--dimensional
grid point number. The total grid point number is $M^3$. 

The PM algorithm computation steps can be summarized as follows:
\begin{itemize}
\item[$\bullet$] given particle coordinates, field source values are assigned 
to the grid points, using some interpolation scheme;
\item[$\bullet$] field equations are solved on the grid and then forces 
     are computed on the grid;
\item[$\bullet$]  the force components are interpolated back to the particles 
and velocities and positions are updated.
\end{itemize}

In the first step each particle distributes its mass to $N_g$ grid points
according to a polinomial interpolation. 
The most popular are the {\it Cloud in Cell} (CIC) and the {\it Triangular
Shaped Cloud} (TSC) schemes.
The first one performs a linear interpolation and scatters the 
particles to the 8 nearest grid points. The
second one (a quadratic interpolation) to the 27 nearest grid points.
In general, $N_g=(n+1)^3$, where n is 
the order of the interpolation scheme.
In particular, in our simulations we use the CIC scheme and the
same number of particles and grid points ($N = M^3$).

In the second step the field equations on the grid is solved. The fastest way 
is to perform a Fast Fourier Transform (FFT) which requires an operation 
number of the order $O(M^3\log M^3)$. The force can be easily computed in 
Fourier domain and then on the grid points, after antitransformation.

In the third part of PM the force field sampled on the grid is interpolated back
to particles. To reduce the errors, the interpolation scheme must be the 
same of density assigment (Hockney \& Eastwood, 1981;
Birdsall \& Langdon, 1985). Finally, positions and 
velocities of the particles are updated. 

The deposition of field-source values on the grid points and the interpolation 
of the force back to the particles are the most time consuming part of codes 
based on PM model: more than $60\%$ of the computation time can be required 
for these steps. 

Furthermore, when the system under analysis reaches strong nonlinear regimes, 
in few time steps and even starting with homogeneously distributed particles, 
it is possible to reach a state with highly clustered particles in 
very few cells, almost uniform distribution of particle in some 
fraction of the cells and no particles in the remaining ones.

The first and the third part have $O(N)$ operation number, whereas the second
one has $O(M^3 \log M^3)$ operation number. Since the particle number N is 
proportional to $M^3$ (typically $N=M^3$ or $N=(M/2)^3$), the total operation
number is $O(N \log N)$.

\section{Parallelization Strategy}

In PM there are two types of data: the particle data, i.e. position and 
velocity, which are stored in one--dimensional arrays of $N$ elements, 
and the grid data, i.e. density and force, which are stored in 
three--dimensional arrays $M\times M\times M$ in size.  Since 
individual particles can change cell under the action of gravitational 
force, parallelization is a real challenge on parallel machine and 
must take into account the machine architecture. 

\begin{figure}
\vskip -2cm
\epsfxsize=\hsize
\epsfbox{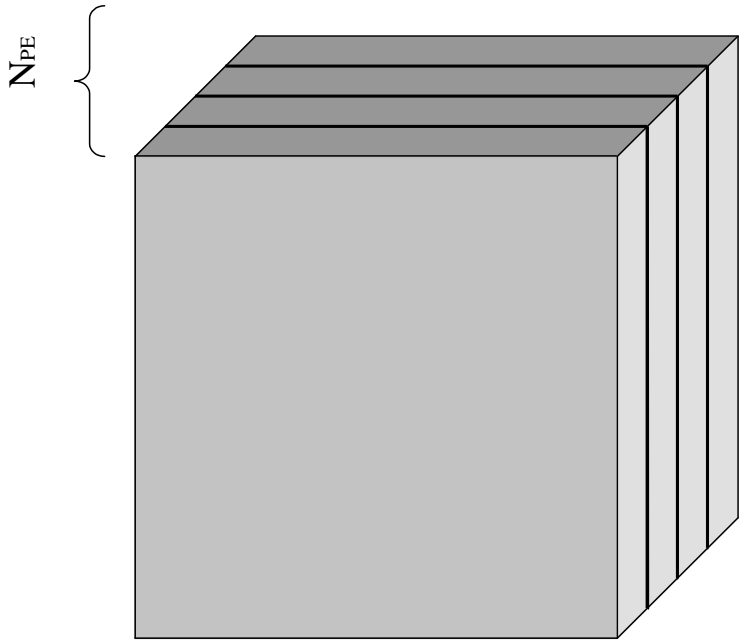}
\vskip -12cm
\caption{Each 3-dimensional array is divided in $N_{PE}$ blocks with respect to
         the third dimension and each processor allocates one block. The first
	 and second dimension are local, while the third is distributed among
	 processors. The grid points are not shown. The figure shows
         a 4 processor case.}
\label{grid1}
\end{figure}
In a naive approach, it is possible to perform a static work distribution 
taking as reference the particles or the grid points. In the first case, 
the machine work is well balanced (each PE has the same number of
 particles), but many remote communications are required, when the 
gravitational system is strongly non--linear. In the CIC scheme, for 
instance, each particle scatters its mass to 8 nearest grid points 
and therefore can require up to 16 remote communications. Due to 
this overhead, the code is inefficient.

In the second approach the grid points are uniformly distributed to 
the processors and each processor performs all computational steps 
for the particles which belong to its grid points. This 
implementation is well balanced if particles are unclustered and 
uniformly distributed in the simulation box, but becomes quickly 
unbalanced as the particles start to cluster.

Since gravitational force has to be studied in non--linear regime, 
these naive approaches are unsuitable.

We approach the problem by a suitable block decomposition of data.
In this scheme
each processor allocates $N/N_{PE}$ particles and $M\times M\times (M/N_{PE})$ 
grid points, where $N_{PE}$ is the number of the processors (see
 figure~\ref{grid1}). In this 
way the first and second dimensions of grid arrays are local while the 
third one and the particles are distributed.
A dynamic work distribution is then implemented to treat this block data.
In this case we will see that the implementation is well balanced,
scalable and efficient.

We describe now the main strategies followed to optimize the 
parallelization on Cray T3E, which are atomic update, agenda 
parallelism and data transfer rate.

\subsection {Atomic Update}
\label{atom}

Parallel codes need sometimes to lock a block data to perform 
atomic operations. For example a block data updated by more processors 
must be updated by the processors in turn. Before a processor updates 
the block data, it must lock them. If another processor wants to 
update the same data and finds them locked, it has to wait for 
their unlocking. As data are unlocked by the first processor, the 
second can proceed to lock and to process the block data.

The {\it shmem} library on Cray (CRAY) contains a routine to 
implement the locking:
{\verb+shmem_swap+} routine performs an atomic swap among two data
on two different processors.

The {\verb+shmem_swap+} prototype is

\vskip 0.7 truecm
\hskip 2 truecm \verb+integer  ires, value, pe, target+

\vskip 0.2 truecm
\hskip 2 truecm \verb+ires = shmem_swap (target, value, pe)+
\vskip 0.7 truecm

This routine puts in variable {\verb+target+} of processor {\verb+pe+} 
the parameter {\verb+value+} and gets in local variable {\verb+ires+} the
previous value of {\verb+target+}.
The following scheme realizes the lock--update:

\vskip 0.7 truecm
\hskip 2 truecm \verb+real  a(N)+

\hskip 2 truecm \verb+integer  ires, value, pe, target+

\hskip 2 truecm \verb+integer  lock+

\vskip 0.2 truecm
\hskip 2 truecm \verb+lock = 0+

\hskip 2 truecm \verb+call barrier()+

\hskip 2 truecm \verb+ires = 1+

\hskip 2 truecm \verb+do while (ires == 1)+

\hskip 2 truecm \verb+    ires = shmem_swap(lock, 1, 0)+

\hskip 2 truecm \verb+enddo+

\hskip 2 truecm \verb+call sub1(a)+

\hskip 2 truecm \verb+ires = shmem_swap(lock, 0, 0)+
\vskip 0.7 truecm 

{\verb+sub1(a)+} is a routine which updates the block-data {\verb+a+}.
To update the block--data {\verb+a+}, the variable {\verb+lock+} on 
$P\!E$ $0$ behaves as a semaphore. If {\verb+lock = 0+} 
or {\verb+lock = 1+} block data are unlocked or locked, respectively. 
Initially its state is set to unlock. The barrier allows all processors 
to know the initial state of block data. {\verb+ires+} contains the 
value of {\verb+lock+}.

To update {\verb+a+}, a processor has to wait {\verb+lock = 0+}. 
After the updating the processor set {\verb+lock = 0+} and allows the 
update of another processor. In this way {\verb+call sub1(a)+} is 
executed in turn by the processors.

\subsection {Agenda Parallelism}

Agenda parallelism is a well known parallel paradigm (Carriero \& Gelernter,
 1990). The work is 
subdivided in an agenda of subworks, $N_{SW}$ in number. Each 
subwork can be marked to notify if it has been performed or not. 
Each processor scrolls the agenda, takes an unmarked subwork and marks 
it. This prevents that another processor performs the same subwork. 
We use the locking procedure to implement the agenda paralellelism, 
through a lock--array of size $N_{SW}$, initially set to 0.
When the work has to be performed, a processor scrolls the lock--array 
of PE 0 and executes any free subwork according to the scheme below:

\vskip 0.7 truecm
\hskip 2 truecm \verb+integer  i, ires, value, pe, target+

\hskip 2 truecm \verb+integer  lock(NSW)+

\vskip 0.2 truecm
\hskip 2 truecm \verb+lock = 0+

\hskip 2 truecm \verb+call barrier()+

\hskip 2 truecm \verb+do i = 1, NSW+

\hskip 2 truecm \verb+    ires = shmem_swap(lock(i),1,0)+

\hskip 2 truecm \verb+    if (ires == 0) call subwork(i)+

\hskip 2 truecm \verb+enddo+

\hskip 2 truecm \verb+call barrier()+
\vskip 0.7 truecm 

\noindent
where {\verb+subwork(i)+} routine performs the $i^{th}$ subwork. 

Agenda parallelism is a good paradigm as the work may be divided 
in subwork with very different computing time. A processor 
performing computationally expensive subworks takes charge of few 
subworks, whereas a processor performing computationally less expensive 
subworks takes charge of many subworks. This procedure is 
self--balancing and each processor spends about the same time to 
complete the whole work. Therefore the agenda parallelism is a good 
approach to implement a dynamic work distribution.

\subsection {Data Transfer Rate}

Parallel computing requires a good data transfer rate among processors.
Cray's {\em shmem} library provides some routines to transfer
data between processors asynchronously. In particular, we use {\verb+shmem_get+} 
and {\verb+shmem_put+} routines which get/put data from/in a processor. 
This routines allow a very fast transfer rate, if many data are transfered 
simultaneously. If the data to transfer are few the transfer rate is 
worse: the transfer time is led by latency time and is independent of 
the data number.

In table \ref{tab1} it is reported the transfer 
time between 2 processors obtained when {\verb+shmem_get+} and {\verb+shmem_put+} 
routines are used.

\begin{table}[h,t]
\centerline{
\begin{tabular} {l cc}
    \hline
    N & t(get) & t(put) \\
    \hline
    $1$		& $9.30~10^{-6}$  & $8.80~10^{-6}$  \\
    $4$		& $1.00~10^{-5}$  & $1.00~10^{-5}$  \\
    $16$	& $1.05~10^{-5}$  & $1.00~10^{-5}$  \\
    $64$	& $1.15~10^{-5}$  & $1.15~10^{-5}$  \\
    $256$	& $1.60~10^{-5}$  & $1.70~10^{-5}$  \\
    $1$ K	& $3.60~10^{-5}$  & $3.60~10^{-5}$  \\
    $4$ K	& $1.06~10^{-4}$  & $1.08~10^{-4}$  \\
    $16$ K	& $4.00~10^{-4}$  & $3.94~10^{-4}$  \\
    $64$ K	& $1.56~10^{-3}$  & $1.54~10^{-3}$  \\
    $256$ K	& $6.20~10^{-3}$  & $6.24~10^{-3}$  \\
    $1$	M	& $2.50~10^{-2}$  & $2.50~10^{-2}$  \\
    $4$	M	& $1.00~10^{-1}$  & $1.00~10^{-1}$  \\
    \hline\\
  \end{tabular}}
  \caption{Transfer time for  {\em shmem\_get} (t(get)) and {\em shmem\_get} (t(put)) routines:
             {\it N} is the number of transfered data and
             {\it t} is the transfer time in second. 4 processors 
	     are in use and the communications
	     are between processor 0 and processor 3.}
  \label{tab1} 
\end{table}

The latency time is very important until when 100 data are transfered
simultaneously. Therefore, as the data are few, it is important to transfer
them by means of one only instruction. It is better to group data
in a unique structure and to transfer this structure, rather then data
separately. Transfering $N$ data with $N$ operations requires about $N$
times the time to transfer all the data simultaneously.

The implementation of density computation and force interpolation
could require many remote--load operations, therefore it is 
important to reduce their number.
 
We introduce a new derived type, which contains all the particle information: 
position and velocity. We call it {\it part3D} and it is described as:
\vskip 0.7 truecm
\hskip 2 truecm \verb+TYPE   part3D+

\vskip 0.2 truecm
\hskip 2 truecm \verb+real     ::  x, y, z+

\hskip 2 truecm \verb+real     ::  vx, vy, vz+
\vskip 0.7 truecm

In this way all 6 data of a particle are in adjacent memory
locations and can be loaded by means of
one remote-load operation. For example, if we need all data of
particle {\verb+j+} of processor {\verb+pe+}, we can use
\vskip 0.2 truecm
\hskip 2 truecm \verb+call shmem_get(XLOC, X(j), 6, pe)+,

\vskip 0.2 truecm 
\noindent 
where {\verb+XLOC+} is a local variable of TYPE {\it part3D} and {\verb+X+} is
a one--dimensional array of TYPE {\it part3D} and size $N$.

\section {Parallel Implementation}
\subsection {Density Computation}

Density computation is the first step of PM. The particles 
scatter their mass to the nearest grid points according to a polinomial 
interpolation.

It is not an easy task to perform a well balanced parallel implementation 
of density
computation in PM algorithm. There are two types of data structures. The first
one is related to particles and is organized in N--element vectors.
Particle positions and velocities are represented by this data type.
3--dimensional arrays are used to represent values on
the grid, as density and force fields.
Particles move with respect to the grid and can cluster around few grid points.
This behaviour is a strong source of load unbalancing in a naive
parallel implementation.

\begin{figure}
\vskip -2cm
\epsfxsize=\hsize
\epsfbox{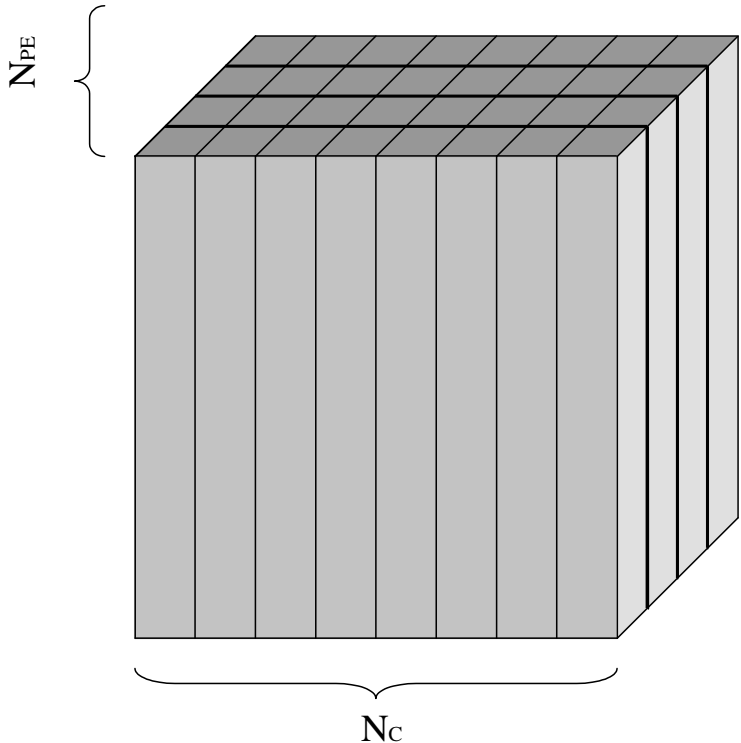}
\vskip -12cm
\caption{ The $N_{PE}$ local array (each of dimension 
          $M\times M\times (M/N_{PE})$) are subdivided in $N_C$ chunks each.
	  Thus the global 3-D array consists of $N_{PE}\times N_C$ chunks.
          In the figure each local array is divided in 8 chuncks.}
\label{grid2}
\end{figure}

To avoid load unbalancing we use a dynamic work distribution
through agenda parallelism (Carriero \& Gelernter, 1990).
We define tasks by dividing the 3--dimensional density arrays allocated
to each processors
($M\times M\times (M/N_{PE})$) in $N_C$ 
chunks of dimension $M\times(M/N_C)\times(M/N_{PE})$ 
(see figure~\ref{grid2}).

Particles are locally sorted so that they are grouped according to the
global chunks they belong to. A local vector contains 
the information on the address of the first particle and on the particle
number for each chunk. In this way for each processor the particles of a chunk
are stored in adjacent memory location and can be remote--loaded together in one 
operation.

To compute the density, each processor selects a task which has not yet been 
executed and locks it. 
Since particles have been ordered 
according to the chunks they belong to, this processor can collect all the
particles 
which give a contribution performing $N_{PE}$ remote--load operations and then 
compute the contribution to the grid density. 

As the task is executed, the processor releases the chunk and selects another 
one from the agenda and then repeats the same operation. This goes on until all
tasks are executed.

In this way each processor manages about the same number of
particle and the load is well balanced. 
If particles are uniformly distributed, the chunks have
about the same number of particles and the processors
work the same number of chunks with a well balanced load.
If particles are strongly clustered, some chunks have many
particles and some few. A processor that grabs such chunks
with many particles, will spend much more time to perform computation
and will take few chunks. On the other hand, a processor which grabs
chunks with few particles, will spend a shorter time
to perform related computations and will take many chunks.
If the total chunk number is much greater than $N_{PE}$,
each processor will treat about the same number of particles and
the load will be well balanced again.

This implementation is dynamic: the processors do not
perform the same work at each time--step and the computational load adapts 
itself to the particle distribution.

Since particles distribute their mass to the nearest grid points, a particle 
near the edge of a chunk updates the
grid points of its chunk and of the next chunks (up to a maximum number of 3)
and a grid point at the edge of a chunk
is updated by particles of its chunk and of nearest chunks
(up to a maximum number of 3). In this way more processors
can update simultaneously the same grid point. 

If this happens, some contribution to density field could be lost.
To avoid this we perform an atomic updating of
density array (see section \ref{atom}).
The processors use a local auxiliary chunk and
scatter the particles on it. It consists of a chunk of the
global density array and of the boundary grid points which the
chunk particles can update.
As the chunk computation is performed, the processor locks in sequence
all the chunks of the global density array, which can be updated by the
chunk particles, and 
updates them by means of the values stored in
the auxiliary one.
In this way another processor cannot update the same
grid points. Finally it unlocks the chunks and makes them available to the
other processors.

\subsection {Force Interpolation}

Force interpolation is the third step of PM algorithm.
In this step the force on each particle and then the particle 
velocity are computed.
It is the inverse operation of density computation, and
we apply the same parallel implementation. 

However there are two main differences.
First, the force arrays must not be updated during force
interpolation. Therefore the lock
of atomic updating is no more required. 
Second, this operation updates the particle velocity, and the
three velocity components must be loaded, updated and stored.
By means of derived type {\it part3D} we load velocity and
position in one operation. So we do not
add remote operation to load velocity. However, to store
the updated velocity we must perform a remote--store operation.
Therefore force interpolation requires two times more communications 
than the density computation.

\subsection {FFT}\label{FFTsec}

Our algorithm solves the field equations by {\it Fast Fourier Transform}
(FFT). There are parallel 3--dimensional FFT routines
in Cray's {\it scilib} library. They operate on 3D array distributed to
the processors. They are very fast but require a lot of workspace memory.
For example, the real--to--complex and complex--to--real FFTs applied
to a 3D array of $M\times M\times M$ size require a complex workspace
of $2\times M\times M\times M$ size.

We have implemented a 3D--FFT for distributed arrays, which is
a bit slower than Cray's implementation, but requires a smaller
workspace.

An analitic three--dimensional Fourier transform (FT) is equivalent to
3 one--dimensional ones,
$$
    \int f({\bf x}) e^{-i{\bf k}\cdot{\bf x}}\, d{\bf x} =
	  \int dz\, e^{-ik_z z} \left( \int dy\, e^{-ik_y y}
	  \left( \int dx \, f(x, y, z) e^{-ik_x x} \right) \right).
$$

Each transformation is performed with respect to one of the 3 variables x, y, z.
The first one
performs 1D--FFTs on column vectors along the array x axis.
The second and the third one perform them on column vectors
along the y and z axis.

In our data distribution the 3D array are distributed between the
processors with respect to the third dimension, whereas the first and
the second one are local. Therefore, each processor performs
the 1D--FFT on all its column
vectors along the x and y axes and the computation is totally local.
The third dimension is distributed and the column vectors along
the z axis are distributed between all processors.
To perform it, we assign the same number of columns to processors.
As a processor computes a Fourier transform for  an assigned column,
collects it from all
processor by {\verb+shmem_get+} routine and stores it in a local vector.
Then the processor performs the 1D--FFT on the local vector and
scatters it to the processors by {\verb+shmem_put+} routine and 
stores it on the distributed array.

To perform the 1D--FFTs we use the routines of {\it scilib} library,
which work on local vectors. To execute them we need a real workspace
of $8\times M$ size per processor. Furthermore we need a complex
auxiliary vector of size $M$ per processor.
So our 3D--FFT implementation requires only a real workspace of size
$10 \times M \times N_{PE}$, which is much less than the $4 \times M
\times M \times M$ real workspace of Cray's implementation.

In table \ref{tab3} we report the execution time for our and Cray's 3D--FFT
implementation on a $256^3$ array.
Our implementation is only slightly slower than Cray's one.
Both routines have a good scalability till $64$ processors. As the
processor number is near to $M$, the number of remote communications
increases and the performance decreases.
Therefore we have implemented a parallel 3D--FFT which requires
much less workspace and is slightly slower than the Cray's one.

\begin{table}[h,b]
\centerline{
   \begin{tabular} {l cccc}
    \hline
    $N_{PE}$		& $16$    & $32$    & $64$    & $128$     \\
    \hline
    {\em Cray}		& $0.83$  & $0.47$  & $0.30$  & $0.22$    \\
    {\em user}		& $1.00$  & $0.51$  & $0.36$  & $0.29$    \\
    {\em Cray/user}  	& $0.83$  & $0.92$  & $0.83$  & $0.76$    \\
  \hline
  \end{tabular}}
  \caption{Computation time for real--to--complex 3D--FFT ruotines
		applied to a $M^3 = 256^3$ 3D array. $N_{PE}$ is the
		processor number. {\em Cray} is the routine of Cray's scilib
		library and {\em user} is our implementation described in
		section~\ref{FFTsec}. {\em Cray/user} is the ratio between the
		computation time of Cray's and our implementations.
                Results are reported for $16$, $32$, $64$
		and $128$ processors. Time is in second.
          }
   \label{tab3} 
\end{table}

\section {Performances}

We have implemented our code on a Cray T3E at CINECA (Bologna).
This T3E has 256 processing
elements (PE) with a 600 MHz DEC Alpha EV5 each. 128
PEs have a 256 MByte memory and the other 128
a 128 MByte memory.

We have compared the result of our implementation with
a serial version of the code (Moscardini, 1990).
The two codes produce results which are the same up
to the machine accuracy.

We performed a cosmological
simulation with $N = M^3 = 256^3$ and $100 h^{-1} Mpc$ box--size ($h$ is 
the Hubble constant in $100 Km\,s^{-1}\,Mpc^{-1}$ unit) to test the 
performances of the parallel implementation with both homogeneous
and clustered particle distributions.

Following Ferrell \& Bertschinger (1994) we
measure the clustering with the homogeneity parameter
$FracOC = N_{oc} / M^3$, where $N_{oc}$ is the number
of mesh cells with at least one particle in them.
$FracOC$ is the fraction of mesh cells with at least one
particle in them.

At the beginning the particle distribution is quite 
homogeous. The gravitational
evolution induces clustering and at the end the particles are clustered.

\begin{table}[h]
\centerline{
   \begin{tabular} {l ccc ccc}
	\hline
    			& & Homogeneous	& & & Clustered	 &	\\
	\hline
				& $32 PE$  & $64 PE$	& $128 PE$ & $32 PE$  & $64 PE$	& $128 PE$	\\
	\hline
	Particle Sorting	& $1.433$  & $0.675$	& $0.326$  & $1.433$  & $0.675$	& $0.326$	\\
	\hline
	Density Computation	& $0.474$  & $0.242$	& $0.135$  & $0.460$  & $0.235$	& $0.130$	\\
	\hline
	FFT			& $2.072$  & $1.464$	& $1.163$  & $2.072$  & $1.464$	& $1.163$	\\
	\hline
	Force Computation	& $0.090$  & $0.048$	& $0.030$  & $0.090$  & $0.048$	& $0.030$	\\
	\hline
	Force Interpolation	& $1.629$  & $0.871$	& $0.460$  & $1.605$  & $0.865$	& $0.457$	\\
	\hline
	Total PM		& $5.698$  & $3.300$	& $2.114$  & $5.662$  & $3.287$	& $2.106$	\\
	\hline
	$FracOC$		& $0.879$  & $0.879$	& $0.879$  & $0.212$  & $0.212$	& $0.212$	\\
	\hline
  \end{tabular}}
  \caption{Computing time for the components of PM in the test simulation.
           Table reports results for $32$, $64$ and $128$ processing 
           elements (PEs) at the begin (homogeneous) and the end (clustered)
           of the simulation. For all test simulations $N = M^3 = 256^3$.
          }
            \label{tab4} 
\end{table}

Table \ref{tab4} shows the computing time for the components of PM at the 
beginning (homogeneous particle distribution)
and at the end (clustered particles). Density Computation and Force 
Interpolation are described above in previous sections. Sorting
is the required time to sort the particle. FFT is the
required time to compute Fast Fourier Transform, 1 forward e 3 
backward. Force Computation collects the routine to compute
the gravitational potential and the 3 force components in
Fourier space. Force Interpolation is performed 3 times
per time--step to compute the 3 force components. Therefore,
Force Interpolation time is about 3 times greater than
Density Computation time. 

\begin{figure}
\vskip -5cm
\epsfxsize=\hsize
\epsfbox{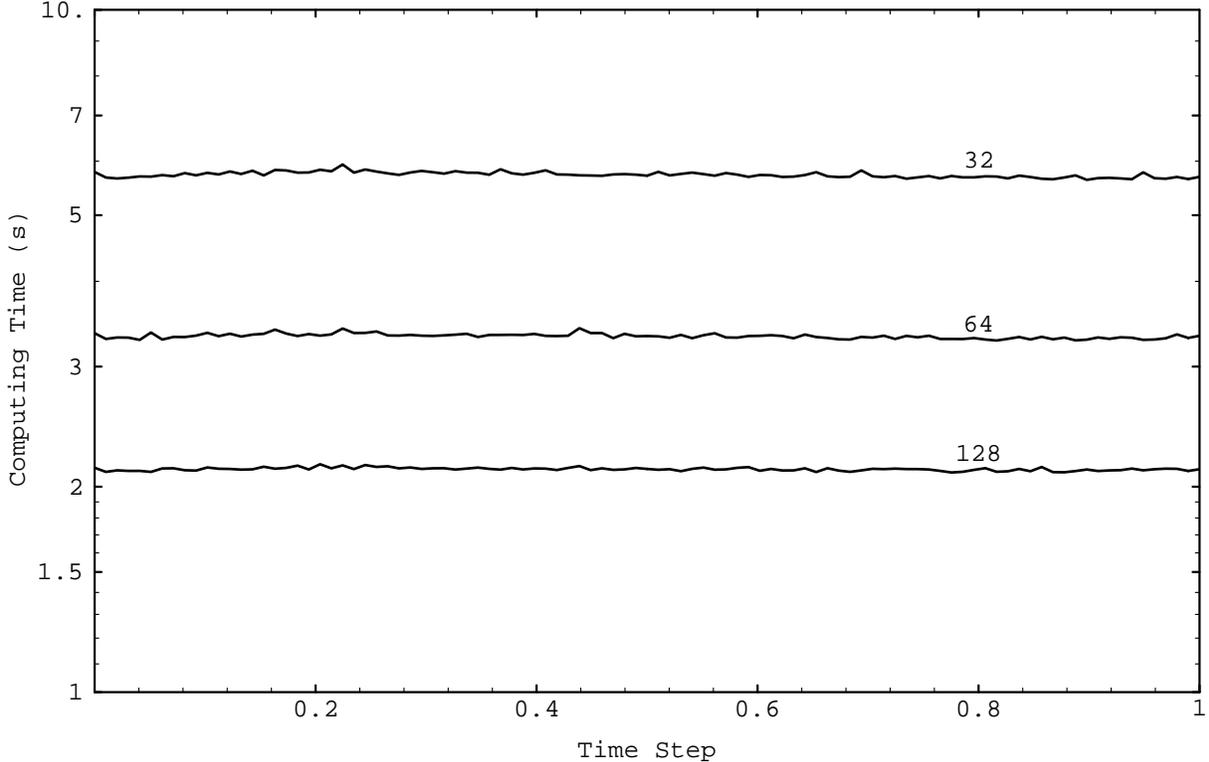}
\vskip -6cm
\caption{Computing time per step (in seconds) with respect to the
          step number. The step is normalized to the total step number.
          We report the results for $32$, $64$ and $128$ processors.}
\label{figure1}
\end{figure}

The balancing properties of the code are shown in 
figure \ref{figure1}, where we plot the computing time per step
of our test simulation for $32$, $64$ and $128$ processors.
The computing time is quite constant for all simulations and
all processor numbers. The small fluctuations are due
only to sorting routine. Therefore our implementation is very well
balanced during the clustering evolution and it is not
affected by the clustering.

\begin{table}[htbp]
\centerline{
   \begin{tabular} {l ccc ccc}
	\hline 
		& & Homogeneous	& & & Clustered	 &	\\
	\hline
		& $32\rightarrow 64$ & $64\rightarrow 128$	& $32\rightarrow 128$	& $32\rightarrow 64$	& $64\rightarrow 128$	& $32\rightarrow 128$ 	\\
	\hline
	Particle Sorting	& $2.12$  & $2.07$  & $4.40$  & $2.12$	& $2.07$  & $4.40$	\\
	\hline
	Density Computation	& $1.96$  & $1.79$  & $3.51$  & $1.96$	& $1.81$  & $3.54$	\\
	\hline
	FFT			& $1.42$  & $1.26$  & $1.78$  & $1.42$	& $1.26$  & $1.78$	\\
	\hline
	Force Computation	& $1.87$  & $1.60$  & $3.0 $  & $1.87$	& $1.60$  & $3.00$	\\
	\hline
	Force Interpolation	& $1.87$  & $1.89$  & $3.54$  & $1.86$	& $1.89$  & $3.51$	\\
	\hline
	Total PM		& $1.73$  & $1.56$  & $2.69$  & $1.72$	& $1.56$  & $2.69$	\\
	\hline\\
  \end{tabular}}
  \caption{Speed--up for the indicated components of PM in the test simulation.
           Table reports results for the indicated processor scaling
           at the begin (homogeneous) and at the end (clustered) of
           the simulation. For all test simulations $N = M^3 = 256^3$.
          }
            \label{tab5} 
\end{table}

The scalable properties are reported in table \ref{tab5} that shows the
speed-up for the components of PM. All components have a speed--up 
near 2, which is the ideal values, and scale very well. Only FFT do not 
scale well, as we have shown above in section~\ref{FFTsec}.
The sorting routine has 
a speed--up greater than 2 because the number of particles per processor
decreases by increasing the processor number. The speed-up of the
whole PM routine is slightly smaller than 2, but this is due to the FFT
routine.

\vskip 0.5cm
\noindent
{\large {\bf {Acknowledgement.}}}This work has been partially supported
by a CNAA grant

\vskip 0.5cm
\noindent

\end{document}